\documentclass[aps,twocolumn]{revtex4}
\usepackage[dvips]{graphics,graphicx}
\usepackage{amsmath}
\begin{document}


\title{Generalization of the Landauer-B\"uttiker theory onto the case of dissipative contacts}
\author{Andrey~R.~Kolovsky}
\affiliation{Kirensky Institute of Physics, Federal Research Center KSC SB RAS, 660036, Krasnoyarsk, Russia}
\affiliation{School of Engineering Physics and Radio Electronics, Siberian Federal University, 660041, Krasnoyarsk, Russia}

\date{\today}
\begin{abstract}
We revisit the problem of two-terminal transport of non-interacting Fermi particles in a mesoscopic device.  First, we generalize the problem by including into consideration relaxation processes in contacts (which are characterized by the contact self-thermalization rate $\gamma$) and then solve it by using the master equation approach.  In the  limit $\gamma\rightarrow0$ the obtained results are shown to reproduce those of the Landauer-B\"uttiker  theory. Thus, the presented analysis  proves analytical correspondence between the Landauer-B\"uttiker and master-equation approaches to quantum transport, -- the problem which resisted solution for decades. 
\end{abstract}
\maketitle

{\em 1.} Quantum transport of identical particles in mesoscopic devices is a vivid topic of non-equilibrium statistical mechanics  \cite{Jin20,Visu22,Uchi22,Senk22,Visu23} which recently got a new impact from fundamentally new experiments with cold atoms in optical potentials \cite{Baro13,Krin17,Corm19}. Theoretically, there are two main approaches for describing the quantum transport in a mesoscopic device. These are the Landauer-B\"uttiker formalism \cite{Land57,Buet85,Datt95} and its further development involving non-equilibrium Green's functions \cite{Meir08,Ramm08,Haug08,Kame11}, and the quantum master equation for the system density matrix \cite{Ajis12,Pros14,Kord15,Land22}. The starting point of both approaches is the same, -- the quantum Hamiltonian which consists of the device Hamiltonian, the two fermionic reservoirs (i.e., the two contacts or leads), and the coupling terms between the system and reservoirs. However, the theories follow different routes. While Green's function methods stay with unitary evolution, the master equation approach deals with non-unitary evolution. Both approaches have their own advantages and disadvantages. The main problem with the master equation approach is the Born-Markov approximation \cite{117,Vadi21}, which leaves outboard the phenomenon of resonant tunneling. The known disadvantage of the second approach is its unitarity which leaves outboard the relaxation and decoherence processes. Although in the past decade these drawbacks of the theories were partially overcome, --  the Green function method was extended to open quantum systems  by using the Keldysh formalism \cite{Sieb16} and the master equation approach was adopted for describing the resonant transmission by relaxing the Markov approximation \cite{122,129} -- the correspondence between the two approaches has not yet been proven. 

In this contribution we  provide the first coherent derivation of the  Landauer-B\"uttiker result by using the master equation approach. We do this by analyzing the paradigm model of quantum transport where two reservoirs of Fermi particles at different chemical potentials are connected by the tight-binding chain of the length $L$. The crucial point of our theory is that we explicitly take into account relaxation processes in the contacts. These processes are characterized by the relaxation rate $\gamma$ which is often referred to as the thermalization or self-thermalization rate. We mention that, although in reality the constant $\gamma$ does depend on temperature, in the  theory we consider these two parameters to be independent of each other.  It is an open problem in the field of many-body Quantum Chaos to relate the rate of self-thermalization of a non-integrable many-body quantum system to its temperature/energy \cite{Nand15,Ales16,Borg16,115}.

{\em 2.} Putting the discussed transport  model on the formal basis we have 
\begin{equation}
\label{Ham_tot}
\widehat{{\cal H}}=\widehat{{\cal H}}_{\rm s}+\sum_{j=1,L}\left(
\widehat{{\cal H}}_{{\rm r},j}+\widehat{{\cal H}}_{{\rm c},j}\right) .
\end{equation}
Here, the subindex '${\rm s}$' stands for the chain (the system), the subindex '${\rm r}$' for reservoirs/contacts, the subindex '${\rm c}$' for the coupling between the system and contacts, and the index $j=1$ and $j=L$ corresponds to the left and right contact, respectively.  In what follows we consider a semi-microscopic model of the contacts/reservoirs given by the tight-binding ring of the size $M$  which eventually tends to infinity. Then the explicit form of the Hamiltonians in Eq.~(\ref{Ham_tot}) is the following
\begin{eqnarray}
\label{Ham_sys}
\widehat{{\cal H}}_{\rm s}=\sum_{\ell=1}^{L}\delta \hat{c}_{\ell}^{\dagger}\hat{c}_{\ell}
-\frac{J}{2}\sum_{\ell=1}^{L-1}\hat{c}_{\ell+1}^{\dagger}\hat{c}_{\ell} +{\rm h.c.} \;,\\
\label{Ham_res}
\widehat{{\cal H}}_{\rm r}=
-J\sum_{k=1}^{M}\cos\left(\frac{2\pi k}{M}\right) \hat{b}_{k}^{\dagger}\hat{b}_{k} \;,\\
\label{Ham_cou}
\widehat{{\cal H}}_{{\rm c}, j}=
-\frac{\epsilon}{2\sqrt{M}}\sum_{k=1}^M\hat{c}_{j}^{\dagger}\hat{b}_{k} +{\rm h.c.} \;.
\end{eqnarray}
Notice that we use the quasimomentum basis for rings and the Wannier basis for the chain.  

In the presence of relaxation process in contacts the governing master equation for the total density matrix ${\cal R}(t)$ of the system (\ref{Ham_tot})-(\ref{Ham_cou}) reads 
\begin{equation}
\label{Master_full}
\frac{\partial \widehat{{\cal R}}}{\partial t}=-i[\widehat{{\cal H}}, \widehat{{\cal R}}]+
\gamma\sum_{j=1,L}\left(\widehat{{\cal L}}^{(g)}_{j}+\widehat{{\cal L}}^{(d)}_{j}\right) ,
\end{equation}
where the Lindblad relaxation operators have the form
\begin{eqnarray}
\label{drain}
\widehat{{\cal L}}^{(d)}_{j}=\sum_{k=1}^M\frac{\bar{n}_{k,j}-1}{2}
\left(\hat{b}_{k}^{\dagger}\hat{b}_{k}\widehat{\cal R }-2\hat{b}_{k}\widehat{\cal R }\hat{b}_{k}^{\dagger}
+\widehat{\cal R }\hat{b}_{k}^{\dagger}\hat{b}_{k} \right) ,\\
\label{gain}
\widehat{{\cal L}}^{(g)}_{j}=-\sum_{k=1}^M\frac{\bar{n}_{k,j}}{2}
\left(\hat{b}_{k}\hat{b}_{k}^{\dagger}\widehat{\cal R }-2\hat{b}_{k}^{\dagger}\widehat{\cal R }\hat{b}_{k}
+\widehat{\cal R }\hat{b}_{k}\hat{b}_{k}^{\dagger} \right) ,\\
\label{Fermi}
\bar{n}_{k,j}= \frac{1}{e^{-\beta_{j}[J\cos(2\pi k/M)+\mu_{j}]}+1} \;.
\end{eqnarray}
This form of the relaxation operators enforce the isolated rings ($\epsilon = 0$) to relax to the thermal equilibrium with the relaxation rate $\gamma$.   For the first time Eq.~(\ref{Master_full}) was introduced in Ref.~\cite{Ajis12} where finite size contacts were referred to as the mesoscopic reservoirs which were coupled to supper-reservoirs at thermal equilibrium. In our setting, where we use the limit $M\rightarrow\infty$, we don't need external reservoirs but it is assumed that the self-thermalization dynamics of the contacts to be mimicked by the Lindblad relaxation operators.

{\em 3.}  Since Eq.~(\ref{Master_full})  is quadratic in terms of creation/annihilation operators, we can derive from this master equation  the equation for the single particle density matrix (SPDM) in the closed form. We have
\begin{equation} 
\label{SPDM}
\frac{d\rho}{dt}=-i[H,\rho]-G*\rho +\gamma\rho_0 \;.
\end{equation} 
In this equation $\rho$ is the SPDM of the size $(M+L+M)\times(M+L+M)$, $H$ the single-particle version of the many-body Hamiltonian (\ref{Ham_tot}), $\rho_0$ the diagonal matrix with elements determined by the Fermi distribution (\ref{Fermi}), $G$ the relaxation matrix with elements proportional to $\gamma$,  
\begin{equation} 
\label{gamma}
{G}_{n,m}=\left\{
\begin{array}{lll}
0&,&(n,m)\in L\times L \\
\gamma&,&(n,m)\in M\times M \\
\gamma/2&,&(n,m)\in L\times M, M\times L 
\end{array}
\right. \;,
\end{equation} 
and the sign '$*$' denotes element-by-element product of two matrices. Setting the left-hand-side of the displayed equation to zero we obtain the algebraic equation for the stationary density matrix which we solve numerically. In what follows we use dimensionless parameters where the hopping matrix element $J$ is the energy unit. Thus, $\epsilon=0.1$ or $\gamma=0.1$ imply that the coupling constant and relaxation rate times the Planck constant are one tenth of the hopping energy. 
\begin{figure}[t]
\includegraphics[width=7.0cm,clip]{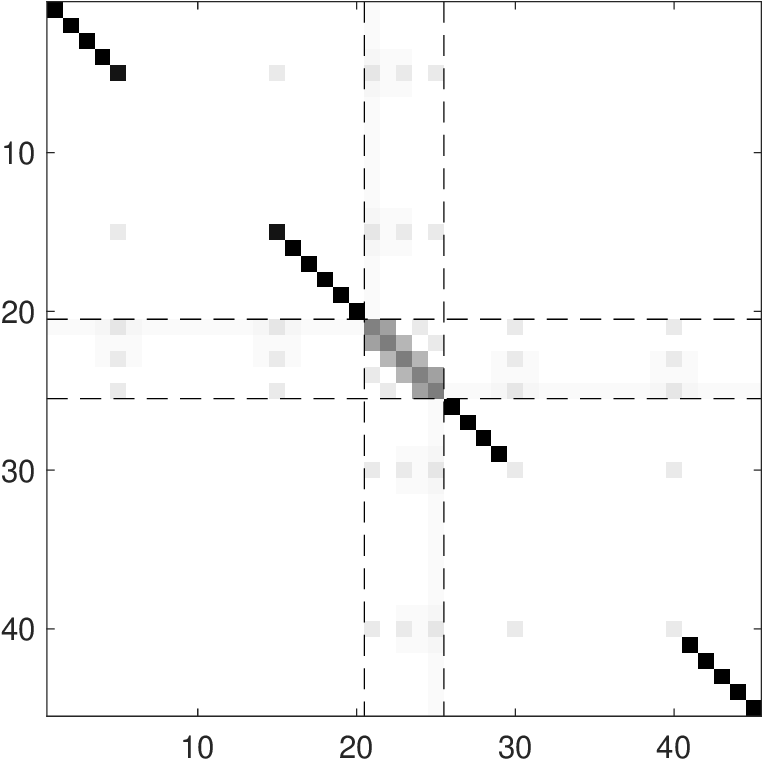}
\caption{Example of the non-equilibrium SPDM. Shown are the absolute values of the matrix elements as the grey-scaled map. System size is $(M,L,M)=(20,5,20)$.}
\label{fig1}
\end{figure} 

As an example, Fig.~\ref{fig1} shows the stationary SPDM for zero temperature and $\mu_L$ and $\mu_R$ slightly above and below $\mu=0$, which would correspond to the half-filling of the isolated contacts. In this figure the central block of the size $5\times 5$ is associated with the chain and the upper and lower blocks of the size $20\times 20$ are associated with the contacts. (Notice that we use the ordering where the zero quasimomentum states $k=M$ is located in the right-lower conner of the block.) One also finds non-zero elements in the off-diagonal blocks of the size $5\times20$ and $20\times5$, which encode entanglement between the system and reservoirs \cite{117}, and strong correlations between the positive and negative quasimomentum states at the ring Fermi energy. Our main interest, however, is the central block which corresponds to the chain SPDM.  Let us discuss this matrix in some more details.

\begin{figure}[b]
\includegraphics[width=8.5cm,clip]{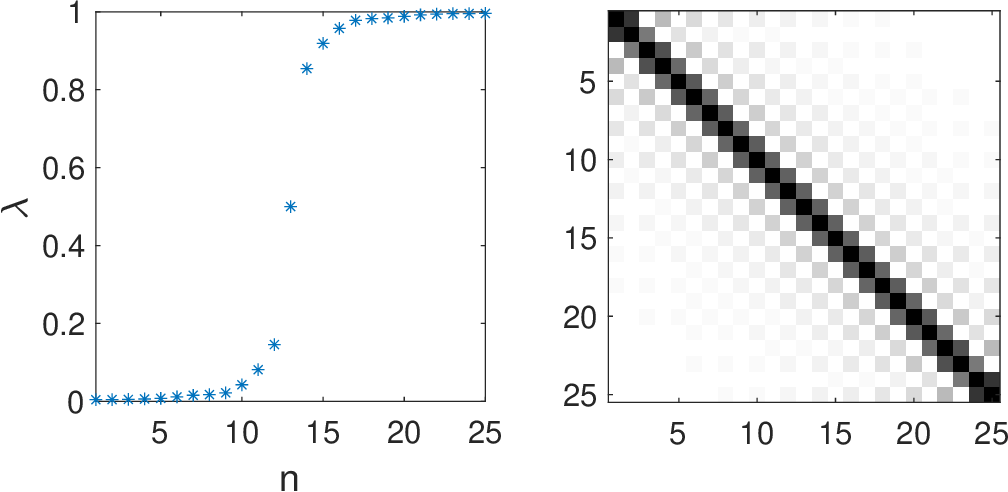}
\caption{The stationary density matrix $\rho_s$ of the chain $L=25$ (right) and its eigenvalues (left). The system parameters are  $\epsilon=0.1$,  $\delta=0$, $\beta=\infty$, $\mu_L=0.05$, $\mu_R=-0.05$, and $\gamma=0.1$.}
\label{fig0}
\end{figure}

The right panel in Fig.~\ref{fig0} shows the density matrix $\rho_s$ for the chain of the length $L=25$, the gate voltage $\delta=0$, zero temperature of the contacts, and the relaxation constant $\gamma=0.1$. The left panel is the matrix spectrum $\lambda_n$, 
\begin{equation} 
\label{SPDM_total}
\rho_s=\sum_{n=1}^L \lambda_n |\psi_n\rangle\langle \psi_n | \;.
\end{equation} 
%
For $\Delta\mu=0$ and $\epsilon\ll 1$ the density matrix eigenstates $|\psi_n\rangle$ with $\lambda_n\approx 1$ and $\lambda_n\approx 0$ are approximately given by the chain eigenstates below and above the Fermi energy. However, for $\Delta\mu\ne0$ the density matrix eigenstates, unlike the chain eigenstates, carry non-zero current. To quantify this difference we introduce the matrix $\tilde{\rho}_s$,
\begin{equation} 
\label{SPDM_delta}
 \tilde{\rho}_s= \frac{\rho_s(\Delta\mu)-\rho_s(\Delta\mu=0)}{\Delta\mu} \;,
\end{equation} 
which is independent of $\Delta \mu$ in the limit $\Delta\mu \rightarrow0$. The matrix $\tilde{\rho}_s$ was analyzed in details in Ref.~\cite{122,129}.
The main conclusion one draws from this analysis is that for any finite $\gamma$ the matrix $\tilde{\rho}_s$ {\em is not} a pure state. 
This is in odd with the common believe that the quantum state of fermionic carriers in the transport channel is given by the Bloch wave with the Fermi quasimomentum.

\begin{figure}[t]
\includegraphics[width=8.0cm,clip]{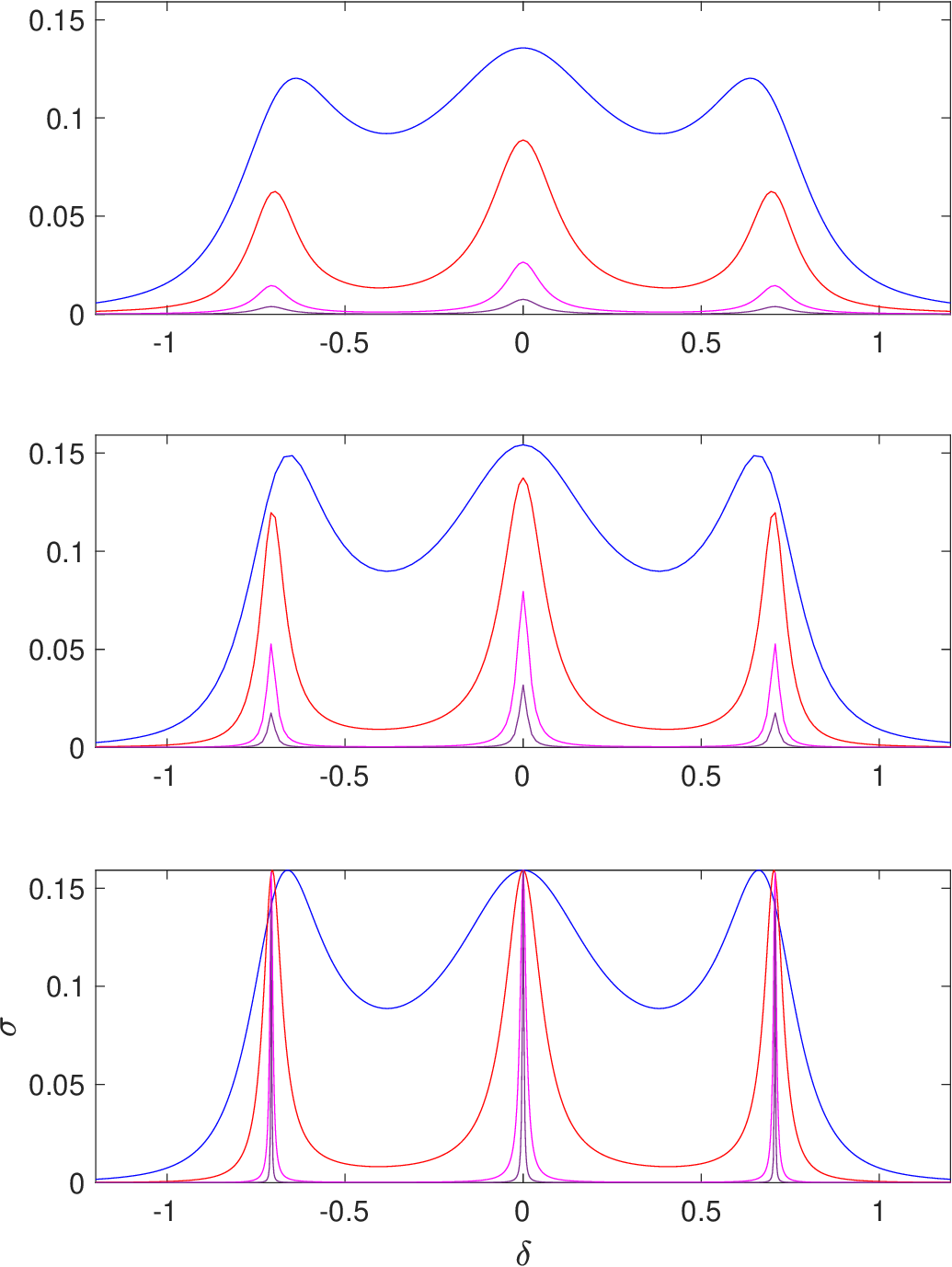}
\caption{Conductance of the chain $L=3$ as the function of the gate voltage for different values of the coupling constant $\epsilon=0.1,0.2,0.5,1.0$ (from bottom to top) and relaxation constant $\gamma=0.1$ (a) and $\gamma=0.02$ (b). The other system parameters are $\beta=\infty$, $\mu=0$, and $M=40$ (a), $M=320$ (b). The panel (c) shows the result according the Landauer-B\"uttiker theory.}
\label{fig2}
\end{figure}

{\em 4.}  Knowing the stationary SPRDM we calculate the system conductance as  
\begin{equation} 
\label{current}
\sigma(\delta)={\rm Tr}[\hat{j}\tilde{\rho}_s]/(L-1)
\end{equation} 
where $\hat{j}$ is the two-diagonal matrix of the single-particle current operator.  Below we calculate conductance Eq.~(\ref{current}) for different system parameters but, first, we discuss the Landauer-B\"uttiker result which will be our  point of reference,

According the Landauer-B\"uttiker theory the chain conductance at zero temperature is given by the  equation 
\begin{equation} 
\label{LB}
\sigma(\delta)=\frac{1}{2\pi} |t|^2 \;,
\end{equation} 
where $1/2\pi$ is the conductance quanta $e^2/h$  in our dimensional units  and $t$ the transmission amplitude which on finds by solving the scattering problem for the plane wave. Since we model the contacts by infinite tight-binding rings, the formulation of the scattering problem  slightly differs from the case of linear tight-binding chain.  Namely, one has to match  the standing-wave  like asymptotic solution in the left ring,
\begin{equation} 
\label{match1}
\psi_m=\left\{
\begin{array}{ccc}
e^{i\kappa m} + r e^{-i\kappa m}   &,& m < 0 \\
e^{-i\kappa m} + r^* e^{i\kappa m} &,& m>0 
\end{array}
\right. \;,
\end{equation} 
to the outgoing-wave asymptotic solution in the right ring, 
\begin{equation} 
\label{match2}
\psi_m=\left\{
\begin{array}{ccc}
t e^{-i\kappa m}  &,& m < 0 \\
t r^* e^{i\kappa m} &,& m>0 
\end{array}
\right. \;,
\end{equation} 
$|r(\kappa)|^2+|t(\kappa)|^2=1$. 
This problem can be easily solved numerically, also in the case where it includes the gate voltage $\delta$ as additional parameter. The lower panel in Fig.~\ref{fig2} shows the conductance Eq.~(\ref{LB}) as the function of $\delta$ for the quasimomentum $\kappa$ given by the Fermi quasimomentum $\kappa_F=\pi/2$ ($E_F=\mu=0$) for four different values of the coupling constant $\epsilon=0.1,0.2,0.5,1.0$. The resonant peaks are clearly seen, where the the chain is perfectly conducting at certain values of $\delta$.

We mention that in the weak coupling limit $\epsilon<1$ the transmission probability $|t|^2$ can found analytically that gives
\begin{equation} 
\label{transmission}
\sigma(\delta) \approx \frac{1}{2\pi}\sum_{n=1}^L 
\frac{\Gamma_n^2}{\Gamma_n^2+(\delta-E_n)^2} \;, \quad  \Gamma_n =\frac{\alpha_n \epsilon^2}{2} \;,
\end{equation} 
where $E_n$ are the eigenvalues of the isolated chain and $\alpha_n$ are determined by the eigenfunctions $\psi_n$ of the isolated chain as 
\begin{equation} 
\label{alpha}
\alpha_n=| \langle \psi_n  | \langle \ell=1\rangle\langle \ell=L | \psi_n\rangle  |  \;.    
\end{equation} 
(Notice that $\sum_n \alpha_n=1$.)  It follows from the displayed equation that the widths of conductance peaks are proportional to $\epsilon^2$ and all peaks have the same height $1/2\pi$.

{\em 5.} We come back to Eq.~(\ref{current}). Fig.~\ref{fig2}(a) shows the conductance $\sigma$  for the relaxation constant $\gamma=0.1$ and the same values of the coupling constant $\epsilon$. The ring size $M$ is determined adoptively by checking that the further increase of $M$ does not affect the result. In full agreement with the analytical prediction of Ref.~\cite{122,129} we see resonant peaks which are locally described by the Lorentzian,
\begin{equation} 
\label{Lorentz}
\sigma(\delta) \sim \epsilon^2\frac{(\gamma/2)}{(\gamma/2)^2+(\delta-E_n)^2} \;. 
\end{equation} 
Thus, now the peak widths are determined by the parameter $\gamma$ while the parameter  $\epsilon$ determines the peaks heights.  It should be stressed that the analytical estimate Eq.~(\ref{Lorentz}) is obtained by using the Born approximation which, in particular, imposes the condition $\epsilon^2/\gamma \ll 1/2\pi$. However, in the numerical analysis we are not bounded by the Born approximation and may consider arbitrary $\epsilon$ and $\gamma$. 

Next we consider the limit of small $\gamma$. Fig.~\ref{fig2}(b) shows the system conductance for five times smaller relaxation constant $\gamma=0.02$.  It is seen in Fig.~\ref{fig2}(b) that the conductance $\sigma=\sigma(\delta)$ converges to the Landauer-B\"uttiker result shown in Fig.~\ref{fig2}(c). Thus, the stationary solution of the master equation for the system density matrix reproduces the Landauer-B\"uttiker result in the limit $\gamma\rightarrow0$. It should be stressed that the convergence to the Landauer-B\"uttiker equation isn't  uniform. For example, for $\epsilon=1$ the conductance calculated on the basis of Eq.~(\ref{SPDM}) coincides with the upper curve in Fig.~\ref{fig2}(c) already for $\gamma=0.01$ while for $\epsilon=0.2$ this requires $\gamma<0.001$.

To conclude the work we give an interpretation of the depicted in Fig.~\ref{fig2} results from the side of Landauer-B\"uttiker equation.  As is mentioned above,  in the weak coupling  limit the system conductance is given by Eq.~(\ref{transmission}).  It is easy to show that Eq.~(\ref{Lorentz}) can be incorporated in the former equation by rewriting it as  
\begin{equation} 
\label{combined}
\sigma(\delta) \approx \frac{1}{2\pi} \sum_{n=1}^L \frac{\Gamma_n(\Gamma_n+\gamma/2)}{(\Gamma_n+\gamma/2)^2+(\delta-E_n)^2} \;. 
\end{equation} 
Eq.~(\ref{combined})  interpolates between the limits $\gamma\ll \epsilon^2$ and $\gamma\gg \epsilon^2$ and  can be viewed as broadening of resonant peaks due to partial decoherence of the carries transporting states in the chain caused by the contact relaxation  dynamics. We compared the estimate (\ref{combined}) with the exact numerical results for different chain length $L$ and different relaxation rates $\gamma$ and found excellent agreement as soon as we restrict ourselves by the weak coupling limit.

{\em 6.} To summarize, we revisit the problem of two-terminal transport of non-interacting Fermi particles across the tight-binding chain by using the semi-microscopic model for the contacts, where we mimic their self-thermalization dynamics by the Lindblad relaxation operators. It is argued that this relaxation  dynamics leads to partial decoherence of the transporting states of fermionic carriers in the chain, which fundamentally  modifies the resonant peak line-shape, see Eq.~(\ref{combined}), which includes the Landauer-B\"uttiker result Eq.~(\ref{LB}) as the particular case $\gamma=0$. 

Proving equivalence between the master equation and Landauer-B\"uttiker approaches (which is the subject of the present contribution) is the first necessary step before addressing the other important physical problems. In fact, conceptional simplicity of the master equation approach allows one to analyze the effect of nonzero $\gamma$ (i.e., the effect of partial decoherence of the transporting states) on the Anderson localization in disordered chains or to study the system conductance in the presence of particle losses, -- the problem which is of particular interest for transport experiments with cold atoms in optical lattices \cite{Baro13}. Besides this, the considered model  is a nice toy model for studying quantum entanglement  between the system and particle reservoirs which, in its turn, controls the validity of the Markov and Born approximations.  The results of these studies will be published elsewhere.

{\em Acknowledge.} The author acknowledges fruitful discussion with D. N. Maksimov and S. V. Aksenov.

 \end{document}